\documentclass[10pt,english]{article}
\usepackage{times}
\usepackage[T1]{fontenc}
\usepackage[latin1]{inputenc}
\usepackage{amsmath}
\usepackage{amssymb}

\makeatletter

\newcommand{\noun}[1]{\textsc{#1}}

\usepackage{slashed}

\usepackage{babel}
\makeatother
\begin{document}

\title{\textbf{Implementing Canonical Seesaw Mechanism in the Exact Solution
of a 3-3-1 Gauge Model without Exotic Electric Charges }}

\author{\noun{ADRIAN} PALCU}

\date{\emph{Department of Theoretical and Computational Physics - West
University of Timi\c{s}oara, V. P\^{a}rvan Ave. 4, RO - 300223 Romania}}

\maketitle
\begin{abstract}
We prove that - even at a low TeV scale - the canonical seesaw mechanism
can be naturally implemented in the exact solution of a particular
3-3-1 gauge model, since a very small alteration $\varepsilon$ in
the parameter matrix of the Higgs sector is taken into account. Therefore,
this new parameter can act as an appropriate mass source for neutrinos,
while - due to the main parameter $a$ - all the previously achieved
results in the exact solution of the model are recovered. Moreover,
this mathematical artifice does separate the boson mass spectrum from
the neutrino mass issue, hence giving more flexibility in tuning the
model. Possible phenomenological results and their implications -
such as dark matter plausible candidates that can occur - are also
briefly discussed. 


\end{abstract}

\section{Introduction}

In a recent paper \cite{key-1}, the author developed a viable method
to generate neutrino masses within the exact solution of a particular
3-3-1 gauge model without exotic electric charges (namely model D
in \cite{key-2}). The procedure relies on the general technique of
exactly solving \cite{key-3} gauge models with high symmetries by
advancing an original parametrization of the Higgs sector. The appealing
feature of our method is that it requires only one free parameter
($a$) to be tuned in order to obtain right predictions for the masses
of the leptons and gauge bosons. These predictions include the values
of the mass splittings for both solar and atmospheric neutrinos. If
these results have to accomodate the present data (\cite{key-4,key-5}
and references therein), the free parameter $a$ must be very small
\cite{key-1}. Once the boson mass spectrum in the model is established
and the right order of magnitude for the neutrino mass splittings
is invoked, the smallness of the parameter leads to a very large order
of magnitude for the overall breaking scale $<\phi>$. In some cases
\cite{key-6} (depending on the choice of the mixing angles) one can
reach a scale very close to the GUT's one $<\phi>\sim10^{16}$GeV.
This could seem quite embarassing when the experimental confirmation
is needed, since very heavy new bosons are predicted by such a large
scale. Therefore, it is very difficult to analyze in detail the whole
resulting phenomenology of the model as long as the breaking scale
is so closely related to the neutrino mass issue. 

In order to avoid the unappealing feature concerning the order of
magnitude for the overall VEV of the model (and hence for the masses
of the new bosons that largely overtake \cite{key-6} the lower limit
supplied by data \cite{key-4}) we intend to turn back to the canonical
seesaw mechanism \cite{key-7} - \cite{key-9} which - we prove in
the following - can be naturally implemented into the exact solution
of the 3-3-1 model of interest even at a low scale about few TeVs.
This can be achieved just by adding a second parameter - much smaller
than the main one - in the Higgs sector. For this purpose we consider
a small alteration ($\varepsilon$) of the parameter matrix $\eta$
of the Higgs sector. This procedure - we prove hereafter - naturally
decouples the breaking scale from the neutrino mass issue and thus
gives more flexibility in tuning the model. The possible dynamical
origin of such a scalar sector remains to be established. 

The paper is organized as follows: in Sec. 2 the implications of the
small alteration ($\varepsilon$) for the exact solution of the model
are analyzed. Then, we recall in Sec. 3 the main features of the canonical
seesaw mechanism that generates neutrino masses with the violation
of the total lepton number at a large scale and try to embed it in
our method by identifying the traditional terms of the seesaw mass
matrix. Some phenomenological results are disscussed in Sec. 4. The
paper ends up with comments on the proposed method and the possible
dark matter candidates that can occur.

\section{The Exact Solution of 3-3-1 Models with Two Free Parameters}

If one is embarrassed by the resulting very high breaking scale \cite{key-1,key-6}
(and very heavy new bosons) in the 3-3-1 model of interest and if
one is ready to deal with two free parameters instead of only one,
we propose a suitable approach. Hoping to get a reasonable scale in
the exact solution \cite{key-1} of the model, one has to adjuste
the $\eta$ parameter matrix with a small amount $\varepsilon$ (let's
call it the fine-tuning parameter) in the following plausible way:

\begin{equation}
\eta^{2}=\left(1-\eta_{0}^{2}\right)diag\left[(a+b)/2+2\varepsilon,1-a+2\varepsilon,(a-b)/2-\varepsilon\right]\label{Eq. 1}\end{equation}
and try to eliminate somehow the parameter $b$. When $\varepsilon\rightarrow0$,
the old parameter matrix (Eq. (10) in Ref. \cite{key-1}) can be recognized.
At the same time, trace condition (Eq. (29) in \cite{key-3}) for
the new $\eta$ has to be accomplished in the manner $\lim_{\varepsilon\rightarrow0}Tr\eta^{2}=\left(1-\eta_{0}^{2}\right)$,
which is obviously true in our new approach. 

Note that the procedure of adding a new small parameter seems not
to affect the previously obtained results in the exact solution of
the model \cite{key-1}, but we still have to check up by calculating
step by step all the masses of the particles. Indeed, when computing
the masses of the bosons in the model, we recover - as one can see
below - the known results \cite{key-1}. When applying the general
procedure (Eq. (55) in \cite{key-3}), the masses of the non-diagonal
bosons become:

\begin{equation}
m_{W}^{2}=m^{2}(a+\varepsilon)\label{Eq. 2}\end{equation}

\begin{equation}
\begin{array}{ccc}
m_{X}^{2}=m^{2}\left(1-\frac{1}{2}a-\frac{1}{2}b+\varepsilon\right), &  & m_{Y}^{2}=m^{2}\left(1-\frac{1}{2}a+\frac{1}{2}b+4\varepsilon\right)\end{array}\label{Eq. 3}\end{equation}
which are - in the limit $\varepsilon\rightarrow0$ - quite the same
values with those obtained in Ref. \cite{key-1} if and only if $\varepsilon$
would not crush the ratio $b/a$ obtained therein from the diagonalization
condition. Evidently, $m^{2}$ replaces $g^{2}\left\langle \phi\right\rangle {}^{2}\left(1-\eta_{0}^{2}\right)/4$
throughout the paper.

The neutral bosons acquire their masses through the following matrix
(applying formula (53) from the general procedure \cite{key-3}): 

\begin{equation}
M^{2}=m^{2}\left|\begin{array}{ccc}
1-\frac{1}{2}a+\frac{1}{2}b+4\varepsilon &  & \frac{-1}{\sqrt{3}\cos\theta}\left(1-\frac{3}{2}a-\frac{1}{2}b\right)\\
\\\frac{-1}{\sqrt{3}\cos\theta}\left(1-\frac{3}{2}a-\frac{1}{2}b\right) &  & \frac{1}{3\cos^{2}\theta}\left(1+\frac{3}{2}a-\frac{3}{2}b\right)\end{array}\right|\label{Eq. 4}\end{equation}
where $\sin^{2}\theta=\frac{4}{3}\sin^{2}\theta_{W}$ \cite{key-1},
since all the charges in the model we deal with have to be preserved. 

Now, one has to check out whether the new parameter $\varepsilon$
alters somehow the ratio $b/a$ between the two main parameters and,
consequently, if the values of the boson masses match the ones finally
obtained in Ref. \cite{key-1}. For this purpose one has to diagonalize
the mass matrix (4), assuming the SM condition between masses $m_{Z}^{2}=m_{W}^{2}/\cos^{2}\theta_{W}$
with $m_{W}^{2}$ given now by Eq. (2). We are surprised to find out
a ratio similar to the ratio in the case when parameter $\varepsilon$
is absent, namely: 

\begin{equation}
b=a\tan^{2}\theta_{W}-\varepsilon\left(\frac{3-4\sin^{2}\theta_{W}}{1-\sin^{2}\theta_{W}}\right)\label{Eq. 5}\end{equation}
It obviously fulfills the required condition $\lim_{\varepsilon\rightarrow0}(b/a)=\tan^{2}\theta_{W}{}$.
Under these circumstances the charged boson masses are: 

\begin{equation}
m_{W}^{2}=m^{2}(a+\varepsilon)\label{Eq. 6}\end{equation}

\begin{equation}
m_{X}^{2}=m^{2}\left[1-a\frac{1}{2\cos^{2}\theta_{W}}+\varepsilon\left(3-\frac{1}{2\cos^{2}\theta_{W}}\right)\right]\label{Eq. 7}\end{equation}

\begin{equation}
m_{Y}^{2}=m^{2}\left[1-\frac{a}{2}\left(1-\tan^{2}\theta_{W}\right)+\varepsilon\left(2+\frac{1}{\cos^{2}\theta_{W}}\right)\right]\label{Eq. 8}\end{equation}
The mass of the Weinberg boson ($Z$) is:

\begin{equation}
m_{Z}^{2}=\frac{m^{2}(a+\varepsilon)}{\cos^{2}\theta_{W}}\label{Eq. 9}\end{equation}
while the new neutral boson ($Z^{\prime}$) develops the following
mass:

\begin{equation}
m_{Z^{\prime}}^{2}=m^{2}\left[1+\frac{1}{3-4\sin^{2}\theta_{W}}-a\left(1+\frac{\tan^{2}\theta_{W}}{3-4\sin^{2}\theta_{W}}\right)+\varepsilon\left(2+\frac{1}{\cos^{2}\theta_{W}}\right)\right]\label{Eq. 10}\end{equation}

We have just obtained the very important following confirmation: when
the new parameter ($\varepsilon$) is sufficiently small it does not
alter the previously obtained structure of the mass spectrum in the
exact solution of the model. At this stage if one desires neutrino
mass, then - obviously - one has to give significance to the small
parameter $\varepsilon$. It could be a very plausible candidate for
playing the role of neutrino mass source if it is embedded in the
Yukawa sector, assuming the same tensor product among Higgs triplets
as in Ref. \cite{key-1}. 

The advantage is that it gives rise to an appropriate seesaw mechanism.
It allows the neutrino mass issue to get a considerable autonomy from
VEV scale. Our new procedure consists of identifying the two terms
in the neutrino sector of the theory corresponding to $a$ and $\varepsilon$
respectively - as it will be outlined in the following section - as
being those responsable for the particular terms of the canonical
seesaw mass matrix \cite{key-7} - \cite{key-9}.

\section{Seesaw Mechanism }

In addition to the above obtained mass spectrum, we have to mention
the new form of $\eta$. It becomes a two-parameter matrix now

\begin{equation}
\eta^{2}=\left(1-\eta_{0}^{2}\right)diag\left[\frac{(a+\varepsilon)}{2\cos^{2}\theta_{W}},1-a+2\varepsilon,\frac{(a+\varepsilon)}{2}(1-\tan^{2}\theta_{W})\right]\label{Eq/ 11}\end{equation}
but its role still remains the same. It determines the correct VEV
alignment in the Higgs sector (consisting of $\phi^{(1)},\phi^{(2)},\phi^{(3)}$)
where $\phi^{(i)}=\eta^{(i)}\phi$ with $i=1,2,3$. Certain cases
can be canceled \cite{key-1} when mapping in a bijective way $(\chi,\eta,\rho)\rightarrow(1,2,3)$
and looking for compatibility with the smallness of the neutrino masses
\cite{key-10} - \cite{key-13}. 

Inspecting under these circumstances the Yukawa sector for leptons 

\begin{equation}
\mathcal{L}_{Y}=G_{\alpha\beta}^{\prime}\bar{f}_{\alpha L}\left(\phi^{(\rho)}e_{\alpha L}^{c}+S{}f_{\beta L}^{c}\right)+G_{\alpha\beta}\varepsilon^{ijk}\left(\bar{f}_{\alpha L}\right)_{i}\left(f_{\beta L}^{c}\right)_{j}\left(\phi^{(\rho)*}\right)_{k}+H.c.\label{Eq. 12}\end{equation}
where $S=\phi^{-1}(\phi^{(\chi)}\otimes\phi^{(\eta)}+\phi^{(\eta)}\otimes\phi^{(\chi)})\sim(\mathbf{1},\mathbf{6},-2/3)$,
one can identify two distinct terms in the neutrino sector of Eq.
(12) - when comparing it to the same situation of the Case I in \cite{key-1}.
Obviously, only Case I out of the three remains, since it is the unique
one that supplies a VEV alignment compatible with small neutrino masses
requirement (see Sec. 4.3 in \cite{key-1}). The two terms are:

\begin{equation}
\mathcal{L}_{Y}^{\nu}=\mathcal{L}_{Y}^{\nu}(a)+\mathcal{L}_{Y}^{\nu}(\varepsilon)\label{Eq. 13}\end{equation}

A natural interpretation of the two terms can occur within the framework
of the canonical seesaw mechanism \cite{key-7} - \cite{key-9} if
one makes the following assumption: $\mathcal{L}_{Y}^{\nu}(\varepsilon)$
corresponds to the Dirac term, and $\mathcal{L}_{Y}^{\nu}(a)$ corresponds
to the righ-handed Majorana term, respectively. Considering the lepton
triplet as $f_{\alpha L}=\left|\begin{array}{ccc}
l_{\alpha L} & \nu_{\alpha L} & (\nu_{\alpha R})^{c}\end{array}\right|^{T}$, this identification leads in the simplest ''one generation case''
to the following neutrino seesaw matrix:

\begin{equation}
M^{D+M}=\left|\begin{array}{ccc}
0 &  & \varepsilon\\
\\\varepsilon &  & 4a\end{array}\right|\frac{\sqrt{1-2\sin^{2}\theta_{W}}}{2\cos^{2}\theta_{W}}<\phi>\label{Eq. 14}\end{equation}
which develops (up to the Yukawa coupling coefficient) the following
mass eigenvalues: 

\begin{equation}
M_{L}^{0}=\left(\frac{\varepsilon^{2}}{a}\right)\frac{\sqrt{1-2\sin^{2}\theta_{W}}}{8\cos^{2}\theta_{W}}<\phi>\label{Eq. 15}\end{equation}
for the left-handed Majorana flavor neutrino, and:

\begin{equation}
M_{R}^{0}=2a\frac{\sqrt{1-2\sin^{2}\theta_{W}}}{\cos^{2}\theta_{W}}<\phi>\label{Eq. 16}\end{equation}
for the very massive seesaw Majorana partner of the left-handed neutrino,
respectively. 

If the neutrino mixing is taken into account as it results from the
Lagrangian (12), then Majorana masses $m_{i}$ for the left-handed
physical neutrinos can be obtained by diagonalizing the symmetric
matrix:

\begin{equation}
M(\nu)=M_{L}^{0}\left|\begin{array}{ccc}
A & D & E\\
D & B & F\\
E & F & C\end{array}\right|\label{Eq. 17}\end{equation}
where the Yukawa couplings in the lepton sector of the model $A=G_{ee}^{\prime}$,
$B=G_{\mu\mu}^{\prime}$, $C=G_{\tau\tau}^{\prime}$, $D=G_{e\mu}^{\prime}$,
$E=G_{e\tau}^{\prime}$, $F=G_{\mu\tau}^{\prime}$ should disappear
by solving an appropriate set of equations for different mixing angles
choices (as is carried out in \cite{key-6}). Assuming the concrete
form of the mass matrix as:

\begin{equation}
M(\nu)=\left(\frac{\varepsilon^{2}}{a}\right)\frac{\sqrt{1-2\sin^{2}\theta_{W}}}{8\cos^{2}\theta_{W}}<\phi>\left|\begin{array}{ccc}
A & D & E\\
D & B & F\\
E & F & C\end{array}\right|\label{Eq. 18}\end{equation}
one can analyze the neutrino mass spectrum just by tuning the parameter
$\varepsilon$ at any breaking scale, since this is determined only
by the parameter $a$. We have just proved that the neutrino mass
spectrum and the VEV issues are decoupled!

\section{Phenomenological Consequences}

If one has to fit all the available data concerning the neutrino mass
splittings \cite{key-5} for different choices of mixing angles, then
the ratio $\varepsilon^{2}/a$ seems to have to be in the range $\sim10^{-15}$
and even smaller \cite{key-6}. 

If one assumes the lower limit $\sim1.5$TeV for the mass of the new
neutral boson in the model (as it is accepted in the data supplied
by \cite{key-4}), then  Eq. (33) in Ref. \cite{key-1} has to be
replaced by a more restrictive condition on the main parameter $a$
which has now to be in the range $a<0.06$. This means the lower limit
of the possible VEV of the model has now increased up to $<\phi>\geq1$TeV.
Consequently, in order to keep the computed neutrino mass squared
differences at their accepted order of magnitude supplied by recent
global analyses $\Delta m_{sol}^{2}\sim8\times10^{-5}$eV$^{2}$ and
$\Delta m_{stm}^{2}\sim2.4\times10^{-3}$eV$^{2}$ \cite{key-5} one
must consider $\varepsilon\sim10^{-8}$ and even smaller. Furthermore,
the masses of the seesaw companions of the physical neutrinos can
be inferred (for the above range of the main parameter $a$) from
a trace condition (using Eq. (16) and coupling coefficients related
to the charged leptons) as: $\sum_{i}m(N_{i})<0.115\cdot\left[m(e)+m(\mu)+m(\tau)\right]$.
That is $\sum_{i}m(N_{i})<216$MeV.

\section{Concluding remarks}

We have presented a plausible method of generating neutrino masses
within the framework of the exact solution of a particular 3-3-1 gauge
model, just by introducing a very small amount $\varepsilon$ into
the parameter matrix of the Higgs sector. This new parameter plays
the role of a second mass source in the model and naturally give rise
to a canonical seesaw mechanism. The advantage of the two-parameter
method is that - apart from the one-parameter case \cite{key-1} -
it does not require a very large breaking scale $<\phi>$ in the model.
That is, neutrino masses can accomodate the experimental data just
by tuning the parameter $\varepsilon$ at any level of the VEV above
TeV scale which is determined only by the main parameter $a$ of the
model. Hence, the masses of the new gauge bosons can come out with
a reasonable order of magnitude that can be tested in the forthcoming
experiments while all the Standard Model phenomenology remains unchanged. 

In addition, our method could give some candidates for the thermally
generated dark matter particles in accordance with general properties
emphasized in recent reviews on this issue (see for example \cite{key-14}
and references therein). Therefore, the Majorana partners of the lightest
physical neutrinos - namely $N{}_{1}$ (if the solution of the model
leads to a normal hierarchy) or $N{}_{3}$ (in the most likely case
of the inverted hierarchy that seems to occur in \cite{key-6}) -
can be taken into account as possible dark matter particles, since
their resulting mass is in the range of MeVs (see MeV fermion dark
matter treated in \cite{key-15}) and they fulfil all the established
conditions \cite{key-14}. Further investigations of the scalar sector
of the model could also reveal some new dark matter candidates (like
in some recent papers \cite{key-16,key-17}) if the self-interacting
Higgs neutral bosons acquire appropriate masses and at the same time
accomplish the stability conditions.


\begin{thebibliography}{10}
\bibitem{key-1}A. Palcu, \emph{Mod. Phys. Lett.} \textbf{A21} (in press), 2006 (hep-th/0601057). 
\bibitem{key-2}W. A. Ponce, J. B. Florez and L. A. Sanchez, \emph{Int. J. Mod. Phys.}
\textbf{A17}, 643 (2002). 
\bibitem{key-3}I. I. Cot\u{a}escu, \emph{Int. J. Mod. Phys.} \textbf{A12}, 1483
(1997).
\bibitem{key-4}S. Eidelman \emph{et. al.}, Particle Data Group, \emph{Phys. Lett.}
\textbf{B592}, 1 (2004). 
\bibitem{key-5}M. Maltoni, T. Schwetz, M. A. Torola and J. W. F. Valle, \emph{New.
J. Phys.} \textbf{6}, 122 (2004). 
\bibitem{key-6}A. Palcu {[}hep-ph / 0605124{]}.
\bibitem{key-7}M. Gell-Mann, P. Ramond and R. Slansky, in \emph{Supergravity} p.
315, edited by F. van Nieuwenhuizen and D. Freedman North Holland
- Amsterdam (1979).
\bibitem{key-8}T. Yanagida, Proc. of the \emph{Workshop on Unified Theory and the
Baryon Number of the Universe}, KEK Japan (1979). 
\bibitem{key-9}R. N. Mohapatra and G. Senjanovic, \emph{Phys. Rev. Lett.} \textbf{44},
912 (1980).
\bibitem{key-10}C. Weinheimer \emph{et al.}, \emph{Nucl. Phys. Proc. Suppl.} \textbf{118},
279 (2003). 
\bibitem{key-11}C. Kraus \emph{et al.}, \emph{Eur. Phys. J.} \textbf{C40}, 447 (2005).
\bibitem{key-12}V. M. Lobashev \emph{et al.}, \emph{Prog. Part. Nucl. Phys.} \textbf{48},
123 (2002). 
\bibitem{key-13}M. Tegmark {[}hep-ph / 0503257{]}. 
\bibitem{key-14}E. A. Baltz, {[}hep-ph / 0412170{]}.
\bibitem{key-15}C. Boehm, D. Hooper, J. Silk and M. Casse, \emph{Phys. Rev. Lett.}
\textbf{92}, 101301 (2004).
\bibitem{key-16}H. N. Long and N. Q. Lan, \emph{Europhys. Lett.} \textbf{64}, 571
(2003).
\bibitem{key-17}S. Filippi, W.A. Ponce and L. A. Sanchez, \emph{Europhys. Lett.} \textbf{73}
(1), 142 (2006).\end{thebibliography}
\end{document}